\documentclass[aps,reprint]{revtex4-1}
\usepackage{amsmath}    
\usepackage{graphicx}  
\usepackage{verbatim}  
\usepackage{color}      
\usepackage{subfigure} 
\usepackage{hyperref}   
\usepackage{xcolor}
\definecolor{TTH-color}{named}{green}

\begin{document}

\title{Thermoelectric radiation detector based on a
superconductor-ferromagnet junction: calorimetric regime}
\author{Subrata Chakraborty }
\email[Correspondence to: ]{subrata.s.chakraborty@jyu.fi}
\author{Tero T. Heikkil{\"a}}
\affiliation{Department of Physics and Nanoscience Center, University of Jyv{\"a}skyl{\"a},
 P.O. Box 35 (YFL), FI-40014 University of Jyv{\"a}skyl{\"a}, Finland}

\date{\today}

\begin{abstract}
We study the use of a thermoelectric junction as a thermal radiation detector
in the calorimetric regime, where single radiation bursts can be
separated in time domain. We focus especially on the case of a large
thermoelectric figure of merit $ZT$ affecting significantly for
example the relevant thermal time scales. This work is motivated
by the use of hybrid superconductor/ferromagnet systems in creating an
unprecedentedly high low-temperature $ZT$ even exceeding unity. Besides
constructing a very general noise model which takes into account cross
correlations between charge and heat noise, we show how the detector
signal can be efficiently multiplexed by the use of resonant
LC circuits giving a fingerprint to each pixel. We show that for
realistic detectors operating at temperatures around 100 to 200 mK,
the energy resolution can be as low as 1 meV. This allows for a broadband
single-photon resolution at photon frequencies of the order or below 1
THz. 

\end{abstract}


\maketitle

\section{Introduction} 
Some of the most sensitive sensors of electromagnetic radiation are
based on using superconducting films absorbing the radiation and
measurement systems converting this process into detectable electronic
signal. The best-studied example of such sensors is the
superconducting transition edge sensor (TES) \cite{TEScitation}, which
has already been used for many types of applications, such as in
security imaging \cite{Luukanen}, materials analysis \cite{Ullom,
  Palosaari} and cosmic microwave background radiation detection
\cite{Hanson, Madhavacheril}. In TES sensors the absorbed radiation
heats the electrons above the critical temperature $T_c$ of 
superconducting films, and results into measurable changes in the
film resistance. This resistance is often read out by utilizing an applied
bias voltage or current \cite{Tero_2006}, fixing the operating point
close to $T_c$, and allowing for additional read-out features such as
electrothermal feedback, and bias-based multiplexing
strategies \cite{Ullom}. However, the presence of bias-induced dissipation also leads to an
overall heating of the system, and increases the thermal noise, thus
reducing the sensitivity. In addition, in multi-pixel systems
fabricating bias lines for each pixel becomes a technological challenge.
Another important sensor in this context
is the kinetic inductance detector (KID) \cite{Grossman, Bluzer, Sergeev, Giazotto, Govenius}
based on the read-out of the kinetic inductance signal in
superconducting microwave resonators. Also KIDs require probe signals
for read-out, resulting into increased dissipation within the
pixels. As the desire in many applications is to further increase the
number of detector pixels \cite{farrah2017far}, such probe-based
sensors become increasingly difficult to operate. 

In general, one would prefer to only have the effect of the coupling
between radiation and the detector in the measured signal, and
therefore to get rid of the probe signal. This desire can be achieved
with a thermoelectric detector (TED) \cite{Jones,
  Varpula,Vechten,Grigory,Tero_bolometer}, where 
the absorption of radiation leads to a temperature difference, which creates a measurable 
thermoelectric current or voltage. In this work we consider such thermoelectric
detectors. They have indeed the advantage of the lack of probe
signals. However, for most systems the thermoelectric effects are very
weak, and therefore sensitivity close to those of TES and KID cannot
be expected. This changed with the discovery of the giant
thermoelectric effect in superconductor-ferromagnet hybrids
\cite{Tero_2014,Kolenda,Tero_2017}, which can in principle be utilized
to create sub-Kelvin thermoelectric heat engines with figures of merit
ZT exceeding even those of the best thermoelectric devices. Hence,
such systems may offer sensitivity rivaling those of TES and KID
sensors, but without the need of probe signals \cite{Tero_bolometer}.

Compared to probe-based sensors, there are also issues with the proper
read-out of the signal, as many existing multiplexing strategies are
based on modulating the probe signal. We address this here by
analyzing in detail the use of the thermoelectric detector in the
calorimetric regime, where radiation arrives at bursts separated by
long times compared to the relevant time scales of the detector. This
is opposite to the bolometric regime analyzed in the earlier work
\cite{Tero_bolometer}. In contrast to many recent works
\cite{Varpula,Vechten} utilizing an ad hoc noise model,
we derive the energy resolution of such a thermoelectric calorimeter
by taking into account all the relevant noise terms, including the
cross-correlation of heat and current noises in the thermoelectric
junction, as required by the linear response theory. As a result, we
obtain the energy resolution and the relevant thermal time scales of the
calorimeter modified by the large ZT. We also analyze the resulting
time-dependent thermoelectrically generated current profile in various parameter regimes.

\begin{figure}[h]
\includegraphics[width=8cm]{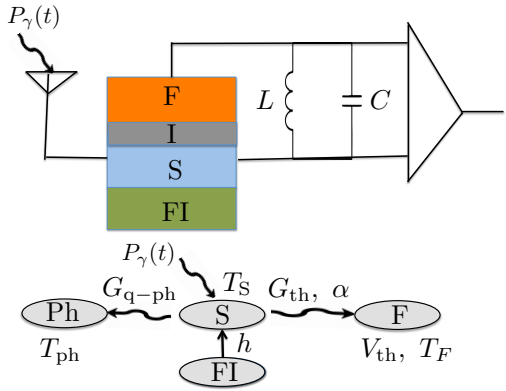}
\caption{\label{fig1} Schematic of the thermoelectric detector based on superconductor (S) and ferromagnetic (F) 
electrode  a spin-filter junction. S is also coupled with ferromagnetic insulator (FI) which provides a spin splitting exchange
field to S. I is an insulating layer and $P_{\gamma}(t)$ is the time dependent power of incident radiation which needs to be detected.
$T_S$, $T_F$ and $T_{\mathrm{ph}}$ are the temperature of the superconducting film,  the temperature of the ferromagnetic
electrode and phonon temperature,  respectively. }
\end{figure}

In this work we consider a single pixel of the thermoelectric
detector based on a superconducting film and the ferromagnetic junction \cite{Tero_bolometer},
as depicted in Fig.~\ref{fig1}. The single pixel of the proposed detector
is built with an element made of a thin film of a superconductor-ferromagnetic
insulator (S-FI) bilayer coupled to superconducting antennas via a clean
(Andreev) contact. Next, the S-FI bilayer is further coupled with a non-superconducting electrode
via a spin filter such as a ferromagnetic metal (F). 
The current injected to the ferromagnetic electrode or the voltage 
between the superconductor-ferromagnet (SF) tunnel junction due to the
incident radiation power  can be detected by a SQUID current amplifier
or a field effect transistor such as HEMT, respectively.
The ferromagnetic insulator (FI) which is in contact with the superconductor 
exerts a spin-splitting exchange field, $h$, into the superconductor. Below certain critical value,
this exchange field does not have a large effect on the
superconducting gap, $\Delta$ \cite{Clogston, Alexander}, but 
it has a major effect on the spin-dependent density of states. 
Another alternative hybrid structure of the detector in this context can be thought of
as superconductor-ferromagnetic insulator-normal metal (S/FI/N). In
that case the FI plays a dual role of providing both the
spin-splitting field via the magnetic proximity effect and the
spin-polarized tunneling required for the thermoelectric effect
\cite{Tero_2017}. However, the tunneling resistance in such junctions
is often larger than in native oxide junctions, leading to somewhat
smaller thermoelectric currents \cite{strambini17,moodera88}.

In what follows, we first study the generic
thermoelectric detectors in the calorimetric regime within the linear response
regime. Then we analyze the energy resolution of the detector
in frequency domain with the idea of optimal filtering. After this we
concentrate on the superconductor-ferromagnet detector and study
the energy resolution vs. bath temperature and the exchange field induced into the
superconductor. Finally, we estimate the practically feasible energy
resolution in such detectors.

\section{Theory}

Here we study the thermoelectric detector in the calorimetric regime,
where the process of relaxation of the detector is much faster than
the arrival of the consecutive incident pulses of energies to the detector.
In what follows  we first consider the heat balance equation of a
generalized thermoelectric detector within linear response assumption
given by \cite{Tero_bolometer}
\begin{equation}
C_h\frac{d\Delta T(t)}{dt} = P_{\gamma}(t)-
G_{\mathrm{th}}^{\mathrm{tot}}\Delta T(t)-\alpha V_{\mathrm{th}}(t),
\label{cal1}
\end{equation}  
where $P_{\gamma}(t)$ is the power of the incident radiation, 
$V_{\mathrm{th}}(t)$ is the time dependent voltage between 
the ferromagnetic electrode and the superconductor, $C_h$ is the heat capacity 
of the absorber and $\alpha$ is the response coefficient for the Peltier heat current.
In Eq.~\eqref{cal1} $\Delta T(t)=T_S(t)-T$ is the change of temperature in the superconductor 
due to the incident power where $T_S(t)$ is the time dependent temperature of the superconductor
and  $T$ is the bath temperature. Here, we consider $T$ equal to the temperature of ferromagnetic
electrode ($T_F$)  and phonon temperature ($T_{\mathrm{ph}}$).
The quantity $G_{\mathrm{th}}^{\mathrm{tot}}=G_{\mathrm{q-ph}}+G_{\mathrm{th}}$ 
represents the total heat conductance of the superconducting film to
the heat bath, 
and $G_{\mathrm{q-ph}}$ and $G_{\mathrm{th}}$ stand
for the heat conductance of the quasiparticles in the superconductor to the phonons
and to the ferromagnetic electrode, respectively. Another possible
(spurious) heat conduction channel could be due to
quasiparticle-magnon scattering, but we disregard it below as it
depends on the microscopic details of the magnets. 
Notably, there has various other 
possibility of spurious heat conduction processes such as via quasiparticle-magnon scattering.  
Equation \eqref{cal1} in the frequency ($\omega$) space gives the
following solution for the change of temperature,
 $\Delta T(\omega)=
\left({P_{\gamma}(\omega)-\alpha V_{\mathrm{th}}(\omega)}\right)/\left({
 G_{\mathrm{th}}^{\mathrm{tot}}} +i\omega C_h\right)$
\footnote{Throughout this work, we use the Fourier transform convention as
$F(t)=\frac{1}{2\pi}\int_{-\infty}^{\infty} d\omega F(\omega)e^{-i\omega t}$}.
Next, within linear response assumption, we consider
the thermoelectric current from the superconductor to the
ferromagnetic electrode \cite{Tero_bolometer},
\begin{equation}
I_{\mathrm{th}}(t) = -\frac{\alpha}{T}\Delta T(t)-G V_{\mathrm{th}}(t) ,
\label{cal3}
\end{equation}
where $G$ is the conductance of the thermoelectric junction.
This thermoelectric current through the thermoelectric junction 
ultimately reaches an amplifier. Disregarding the back-action  noise from
the amplifier and considering the amplifier as a capacitor or an inductor,
the thermal current through the amplifier due to the thermoelectric voltage,
$V_{\mathrm{th}}$, in frequency ($\omega$) space is 
$I_{\mathrm{th}}(\omega) =  V_{\mathrm{th}}(\omega)\left[i\omega C + {1}/({i\omega L})\right]$,
where $C$ and $L$ are the capacitance and inductance of the
detection circuit, respectively. Here $L$ and $C$ can represent
elements that have been designed on purpose for identifying the pixel
(see below).
Next, using Eqs.~\eqref{cal1} and \eqref{cal3}, along with the expression of the
current through the amplifier we obtain the voltage across the thermoelectric junction
$V_{\mathrm{th}}(\omega) = \lambda_V(\omega)P_{\gamma}(\omega)$,
where $\lambda_V(\omega) = {\alpha}/\left[\alpha^2-{Y_{\mathrm{th}}^{\mathrm{tot}}(\omega)Y_{\mathrm{tot}}(\omega)T }\right]$,
$Y_{\mathrm{th}}^{\mathrm{tot}} (\omega)=G_{\mathrm{th}}^{\mathrm{tot}}+ i\omega C_h$ and $Y_{\mathrm{tot}}(\omega)=G +i\omega C
+ 1/(i\omega L)$. The current through the inductor is 
$I_L(\omega) = {V_{\mathrm{th}}(\omega)}/({i\omega L}) =\lambda_I(\omega)P_\gamma(\omega)$,
where $\lambda_I(\omega) = {\lambda_V(\omega)}/({i\omega L})$.
Finally, we obtain the expressions of  $V_{\mathrm{th}}(t)$ and $I_L(t)$,
\begin{eqnarray}
V_{\mathrm{th}}(t) &=& \frac{1}{2\pi}\int_{-\infty}^{\infty} d\omega~ \lambda_V(\omega)P_{\gamma}(\omega)e^{-i\omega t} \label{cal7a} \\
I_L(t) &=& \frac{1}{2\pi}\int_{-\infty}^{\infty} d\omega~ \lambda_I(\omega)P_{\gamma}(\omega)e^{-i\omega t}. \label{cal7} 
\end{eqnarray}
The integrals in Eq.~\eqref{cal7a} and \eqref{cal7} can easily be solved through 
the Cauchy residue theorem, but the general expressions are too long to be presented here.
Rather, below we present some limiting cases.

Next, for the analyses of the various fluctuation processes present in the system,
relevant for predicting  the energy resolution of detection, we consider a Langevin noise circuit model.
We denote $\delta T$, $\delta V$ and $\delta I_L$
as the temperature fluctuation on the absorber, voltage noise across the
capacitor and the current noise across the
inductor. These noises are governed by the charge current noise
$\delta I$ and heat current noise $\delta \dot{Q}_J$ across the
thermoelectric junction, and the heat current noise $\delta \dot{Q}_{\mathrm{q-ph}}$
due to quasiparticle-phonon scattering. These noise terms satisfy the
heat balance equation and the Kirchoff law for the noise terms in 
$\omega$ space \cite{Blanter}, as
\begin{eqnarray}
Y_{\mathrm{th}}^{\mathrm{tot}}(\omega)\delta T(\omega)
&=& \delta \dot{Q}_J (\omega)+\delta \dot{Q}_{\mathrm{q-ph}} (\omega)- \alpha\delta V(\omega) \label{cal11}\\
Y_{\mathrm{tot}}(\omega)\delta V (\omega) &=& \delta I(\omega)
- \frac{\alpha}{T}\delta T(\omega). \label{cal12}
\end{eqnarray} 
Solving Eqs.~\eqref{cal11} and \eqref{cal12}, we can obtain the
voltage noise $\delta V (\omega)$ through the capacitor and 
the current noise through the inductor as $\delta I_L (\omega)= \delta V
(\omega)/(i\omega L)$. We obtain the expressions of the noise
correlations $\left<\delta V(t)\delta V(t^\prime)\right>$ and
$\left<\delta I_L(t)\delta I_L(t^\prime)\right>$, in time domain through 
$4\pi^2\left<\delta V(t)\delta V(t^\prime)\right>
= \int_{-\infty}^{\infty}\int_{-\infty}^{\infty}
d\omega~d{\omega^\prime}~\left<\delta V(\omega)\delta V(\omega^\prime)\right>$
$e^{-i\omega t}e^{-i\omega^\prime t^\prime}$
and  $4\pi^2 \left<\delta I_L(t)\delta I_L(t^\prime)\right>
= \int_{-\infty}^{\infty}\int_{-\infty}^{\infty}
d\omega~d{\omega^\prime}~\left<\delta I_L(\omega)\delta I_L(\omega^\prime)\right>
e^{-i\omega t}e^{-i\omega^\prime t^\prime}$.
To obtain the second order correlations of these noise terms, we consider the intrinsic
correlations of the detector as
\begin{eqnarray}
&& \left<\delta I(\omega)\delta I(\omega^\prime)\right> = 4\pi k_BTG \delta(\omega+\omega^\prime) \nonumber \\
&&\left<\delta \dot{Q}_J(\omega)\delta \dot{Q}_J(\omega^\prime)\right> = 4\pi k_BT^2G_{\mathrm{th}} \delta(\omega+\omega^\prime) \nonumber \\
&& \left<\delta I(\omega)\delta \dot{Q}_J(\omega^\prime)\right> = -4\pi k_BT\alpha \delta(\omega+\omega^\prime) \nonumber \\
&& \left<\delta \dot{Q}_{\mathrm{q-ph}}(\omega)\delta \dot{Q}_{\mathrm{q-ph}}(\omega^\prime)\right> = 4\pi k_BT^2 G_{\mathrm{q-ph}}\delta(\omega+\omega^\prime) \nonumber \\
&& \left<\delta \dot{Q}_{J}(\omega)\delta \dot{Q}_{\mathrm{q-ph}}(\omega^\prime)\right> = 0 \nonumber \\  
&& \left<\delta I(\omega)\delta \dot{Q}_{\mathrm{q-ph}}(\omega^\prime \right> = 0. \nonumber
\end{eqnarray}
Vanishing intrinsic correlations signify that the noises of the corresponding processes are independent.
Finally, we obtain the following simplified expressions for the second order noise correlations as
\begin{eqnarray}
\left<\delta V(t)\delta V(t^\prime)\right>
&=& \frac{1}{2\pi}\left(\frac{4k_BT^2G_{\mathrm{th}}^{\mathrm{tot}}}{ZT}\right)\int_{-\infty}^{\infty}~d\omega~
\left|\lambda_V(\omega)\right|^2 \nonumber \\
&& \times \left[1+(1+ZT)\tau_{th}^2\omega^2\right] e^{i\omega(t-t^\prime)}
\label{cal12c} \\
\left<\delta I_L(t)\delta I_L(t^\prime)\right>
&=& \frac{1}{2\pi}\left(\frac{4k_BT^2G_{\mathrm{th}}^{\mathrm{tot}}}{ZT}\right)\int_{-\infty}^{\infty}~d\omega~
\left|\lambda_I(\omega)\right|^2  \nonumber \\
&& \times \left[1+(1+ZT)\tau_{th}^2\omega^2\right] e^{i\omega(t-t^\prime)}.
\label{cal12d} 
\end{eqnarray} 
In Eqs.~\eqref{cal12c} and \eqref{cal12d} 
$ZT=\alpha^2/(G_{\mathrm{th}}^{\mathrm{tot}}GT-\alpha^2)$ is the thermoelectric figure of merit
and $\tau_{th}=C_h/G_{\mathrm{th}}^{\mathrm{tot}}$ is the thermal relaxation time.
Our theoretical formalism for the current through the inductor and 
the second order noise correlation help us to analyze 
the optimum energy resolution of the thermoelectric detector in the
calorimetric regime. As a result we can find the condition for
single photon detection in the far-infrared regime, as shown below. 

\section{Results}
In this section we discuss the results obtained from the formalism in the previous section.
First we evaluate $I_L(t)$. Next, we obtain the optimal
energy resolution in the calorimetric regime, with the idea of optimal filtering.
This approach helps us analyzing a scheme for multiplexing the read-out.
Finally, we predict the energy resolution in a SF based TED.
 
\subsection{Current through the inductor in the calorimetric regime}
Here we analyze the behavior of the current through the inductor with respect to
time at various circumstances in the calorimetric regime, that is when
$P_{\gamma}(t)=E\delta(t)$. Instead of the current, one could measure the voltage across the capacitor. The results are qualitatively similar, and the intrinsic energy resolution is the same in both cases.

In what follows we denote the charge relaxation time $\tau_{RC}=C/G$ and LC time $\tau_{LC}=\sqrt{LC}$.
For simplicity, we also define the corresponding frequencies by 
$\omega_{th}=1/\tau_{th}$, $\omega_{RC}=1/\tau_{RC}$ and $\omega_{LC}=1/\tau_{LC}$.
\begin{figure}
\includegraphics[width=8cm]{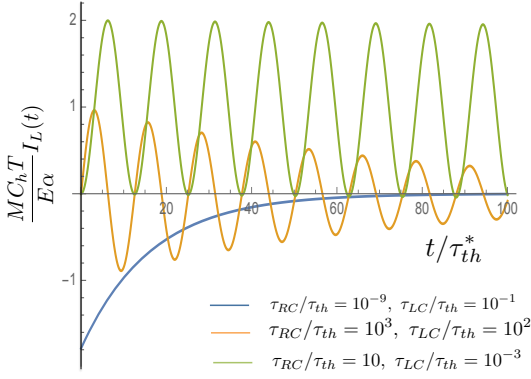}
\caption{\label{fig2} Current $I_L(t)$ through the inductor in
three parameter regimes: fast charge relaxation $\tau_{RC}\ll \tau_{th},~\tau_{LC}$ (blue),
fast thermal relaxation $\tau_{th}\ll \tau_{RC},~\tau_{LC}$ (orange) and 
high resonator frequency $\tau_{LC}\ll \tau_{th},~\tau_{RC}$ (green). In all curves $ZT=1$.
$M$ is a scaling factor which takes different values for blue, orange and green curves as
$10^8$, $10^2$ and $1$, respectively. The time scale  $\tau_{th}^*$ has the 
values $200~\tau_{th}$, $50~\tau_{th}$ and $5\times 10^{-4} ~\tau_{th}$
for blue, orange and green curves respectively.}
\end{figure}
Using Eq.~\eqref{cal7}, first we obtain $I_L(t)$ for finite $t \gg \tau_{RC}$, when the charge relaxation process is fastest,
that is for $\tau_{RC}\ll \tau_{th},~\tau_{LC}$, as 
\begin{subequations}
\begin{equation}
 I_L(t) = \frac{E\alpha}{C_hT (1+ZT)}
\omega_{LC}^2 \omega_{RC}\omega_{th}\phi(t)  \label{cal13} 
\end{equation}
\begin{equation}
 \phi(t) = \exp\left[{ \frac{-\omega_{th}t}{1+ZT} }\right] - 
\exp\left[\frac{-\omega_{LC}^2(1+ZT)t}{\omega_{RC}}  \right].   \label{cal13a}   
\end{equation}
\end{subequations}
Thus in the case of fast charge
relaxation $I_L(t)$ tends to zero for $t\gtrsim \omega_{RL}/(\omega_{RC}\sqrt{1+ZT})^2=LG/(1+ZT)$,
but the initial decay is governed by  the time scale
$(ZT+1)\tau_{th}$. Next, we obtain the expression of $I_L(t)$ for
finite non-zero $t\gg \tau_{th}$, when the thermal relaxation time is the fastest one, that is $\tau_{th}\ll \tau_{RC},~\tau_{LC}$. It is  
\begin{eqnarray}
\hspace{-0.6cm} && I_L(t) = \frac{4E\alpha \omega_{th}Y}{1+ZT}\exp\left[-\frac{\omega_{RC}t}{2(1+ZT)}\right]  
 \sin\left(\frac{Yt}{1+ZT}\right) \label{cal14} \\
\hspace{-0.6cm}&& 2Y = \sqrt{4\omega_{LC}^2(1+ZT)^2-\omega_{RC}^2} ~.
\label{cal14a}
\end{eqnarray} 
In this case $I_L(t)$
is a decaying oscillatory function with a decaying time scale
$2(ZT+1)\tau_{RC}$ if $Y$ is real. On the other hand,
$I_L(t)$ simply decays, if $Y$ is imaginary. Finally, we analyze $I_L(t)$
for $t\gg \tau_{LC}$ for a high resonator frequency $\omega_{LC}$, that is when $\tau_{LC}\ll \tau_{th},~\tau_{RC}$.
In this case 
\begin{eqnarray}
&&\hspace{-7mm} I_L(t) = \left(\frac{E\alpha\omega_{LC}^2}{CC_hLT}\right) 
\left[4e^{-\omega_{RC} t/2}\cos(\omega_{LC}t)- e^{-\omega_{th} t} \right].
\label{cal15}
\end{eqnarray} 
In this case $I_L(t)$  oscillates
with the frequency $\omega_{LC}$. The oscillations decay within the charge relaxation time.
These oscillations are visible especially when the first term dominates, {\it{i.e.}}, $\tau_{RC}\gg \tau_{th}$.
Example time-dependent currents corresponding to these three regimes are shown in Fig.~\ref{fig2}.   

\subsection{Energy resolution in frequency domain with optimal filtering}
In what follows we first optimize the energy resolution using the optimal filtering technique \cite{McCammon2005}. 
From Eqs.~\eqref{cal7a} and \eqref{cal7}
we obtain expressions for the thermoelectric voltage and current
through the inductor as
\begin{eqnarray}
V_{th}(t) &=&\frac{1}{2\pi} \int_{-\infty}^{\infty}~d\omega~E\lambda_V(\omega)e^{-i\omega t}p(\omega), \label{en1}\\
I_L(t) &=&\frac{1}{2\pi} \int_{-\infty}^{\infty}~d\omega~E\lambda_I(\omega)e^{-i\omega t}p(\omega), \label{en2}
\end{eqnarray}
where we have considered $P_{\gamma}(\omega)=Ep(\omega)$
in Eqs.~\eqref{cal7a} and \eqref{cal7}. Next, following Eqs.~\eqref{cal12c}
and \eqref{cal12d} we have the noise correlations for the thermoelectric
voltage and current through the inductor as 
\begin{eqnarray}
\left<\delta V(t)\delta V(t^\prime)\right>
&=& \frac{1}{2\pi}\int_{-\infty}^{\infty}~d\omega~
e_V^2(\omega) e^{-i\omega(t-t^\prime)} \label{en3}\\
\left<\delta I_L(t)\delta I_L(t^\prime)\right>
&=&\frac{1}{2\pi} \int_{-\infty}^{\infty}~d\omega~
e_I^2(\omega) e^{-i\omega (t-t^\prime)}, \label{en4}
\end{eqnarray}
where
\begin{eqnarray}
\hspace{-0.6cm} && e_V^2(\omega)
= \left(\frac{4k_BT^2G_{\mathrm{th}}^{\mathrm{tot}}}{ZT}\right)
\left|\lambda_V(\omega)\right|^2 
 \left[1+(1+ZT)\omega^2\tau_{th}^2 \right] ~ \label{en5}\\
\hspace{-0.6cm} && e_I^2(\omega)
= \left(\frac{4k_BT^2G_{\mathrm{th}}^{\mathrm{tot}}}{ZT}\right)
\left|\lambda_I(\omega)\right|^2 
 \left[1+(1+ZT)\omega^2\tau_{th}^2 \right]. \label{en6}
\end{eqnarray}
From Fig.~\ref{fig2} we can see that when $\tau_{RC}$ is not the shortest time scale,
$I_L(t)$ decays with oscillation as $t$ increases.
Here our aim is to find the best estimate of 
$E$ in the presence of signal noise terms as Eqs.~\eqref{en3} and \eqref{en4}.
Equations \eqref{en3} and \eqref{en4} indicate that the noise terms are correlated at different
times and uncorrelated in frequency space, therefore it is easier to make an analysis in frequency domain.
Now, let us choose a weight function $W(\omega)$ in the frequency domain,
and therefore define the expected values for the signal and the
corresponding noise terms as 
$\left<V_{th}\right> = \frac{1}{2\pi}\int_{-\infty}^{\infty}d\omega~ W(\omega)E\lambda_V(\omega)p(\omega)$,
$\left<I_L\right> =\frac{1}{2\pi} \int_{-\infty}^{\infty}d\omega~ W(\omega)E\lambda_I(\omega)p(\omega) $,
$\left<\delta V^2\right> =\frac{1}{2\pi} \int_{-\infty}^{\infty}d\omega~ 
\left|W(\omega)\right|^2 e_V^2(\omega) $ and
$\left<\delta I_L^2\right> = \frac{1}{2\pi}\int_{-\infty}^{\infty}d\omega~ 
\left|W(\omega)\right|^2 e_I^2(\omega)$.
Next, we can define the energy resolution of a generalized thermoelectric detector in terms of noise fluctuations
and expected signals as $\Delta E = E{\sqrt{\left<\delta V^2\right>}}/{\left<V_{th}\right>}
= E{\sqrt{\left<\delta I_L^2\right>}}/{\left<I_L\right>}$ \cite{McCammon2005}.
At this point, as we desire to have the maximum value of signal to noise ratio, and hence
the minimum energy resolution, we need to search for an optimal filter, that is an optimal $W(\omega)$.
The desired optimal filter can be found out by finding a zero of the functional derivative of $\Delta E$ with respect to 
$W(\omega)$ \cite{McCammon2005}. The optimal filter is 
$W(\omega)=E\lambda_V(-\omega)p(\omega)/e^2_V(\omega)=E\lambda_I(-\omega)p(\omega)/e^2_I(\omega)$.
Using this weight function we obtain an expression of the optimum energy resolution through such an
optimal filter in the calorimetric regime, that is when $P_{\gamma}(t)=E\delta(t)$ 
or equivalently $p(\omega)=1$, given by
\begin{eqnarray}
\Delta E_{\mathrm{opt}}^{(\mathrm{Fil})} &=& NEP\sqrt{\tau_{\mathrm{eff}}}, \label{en12}
\end{eqnarray}
where $NEP^2=4k_BT^2G_{th}^{tot}/ZT$ and $\tau_{\mathrm{eff}}=\tau_{th}\sqrt{1+ZT}$.
In Eq.~\eqref{en12} the effective time constant, $\tau_{\mathrm{eff}}$ is affected by $ZT$. 
This is the generalization of the energy resolution obtained earlier for a thermoelectric detector
in the calorimetric regime with a non-zero $ZT$ \cite{Vechten}.

In the above analysis about energy resolution with the idea of optimal
filtering, we have not considered the added noise in amplification,
described by the low-frequency power spectral density $S_A$. Now, in the measured signal, we consider an effect
due to noise term $\delta I_A(t)$ of a current amplifier, where amplifier noise term is uncorrelated with other intrinsic noise terms of
the detector. We assume the amplifier current noise fluctuation in the calorimetric regime is 
$\left< \delta I_A(t) \delta I_A(t^\prime) \right> = \frac{1}{2\pi} \int_{-\infty}^{\infty} d\omega~ S_A~ e^{-i\omega\left(t-t^\prime\right)}$
Therefore in order to obtain the optimal energy resolution, we design the optimal filter by including the
effect due to amplifier current noise term. As a result, we get the
optimal energy resolution
$\Delta E_{\mathrm{opt}}^{(\mathrm{Fil})} = NEP_{\mathrm{tot}}\sqrt{\tau^{\mathrm{tot}}_{\mathrm{eff}}}$,
where $NEP_{\mathrm{tot}}^2=4k_BT^2G_{th}^{tot}/ZT_{\mathrm{tot}}$ and $\tau^{\mathrm{tot}}_{\mathrm{eff}}=\tau_{th}\sqrt{1+ZT_{\mathrm{tot}}}$. 
Here $ZT_{\mathrm{tot}}^{-1}=ZT^{-1} + ZT_A^{-1}$, where
$1/ZT_A=S_AG_{th}^{tot}/(2k_B \alpha^2)$. The effect of the amplifier can
hence be disregarded if $ZT/ZT_A =  S_A (1+ZT)/(2G k_B T) \ll 1$.

\subsection{Comment about multiplexing}
\begin{figure}
\includegraphics[width=8cm]{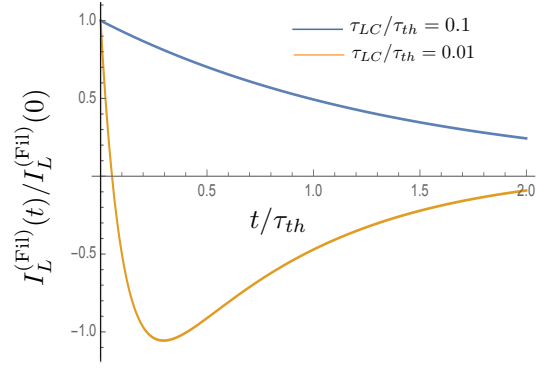}
\caption{\label{fig3} Filtered current $I_L^{\mathrm{(Fil)}}(t)$ through the inductor.
In all curves $\tau_{RC}/\tau_{th}=10^{-3}$ and $ZT_{\mathrm{tot}}=1$. In the blue curve 
$W(\omega)$ we choose the optimal filter.
On the other hand for the orange curve, we consider the $W(\omega)$
is equal to the optimal filter of the blue curve. }
\end{figure}

In this section we consider a practical multiplexing case by doing a numerical experiment.
For this, we define the filtered signal in the calorimetric regime from the previous section as
\begin{eqnarray}
I_L^{\mathrm{(Fil)}}(t) &=&  \frac{E}{2\pi} \int_{-\infty}^{\infty} d\omega~ W(\omega) \lambda_I(\omega) e^{-i\omega t} \label{multiplex1}
\end{eqnarray}
Now, in Eq.~\eqref{multiplex1} if we consider $W(\omega)$ to be the optimal filter as obtained in the previous section, 
then we have
\begin{eqnarray}
&& I_L^{\mathrm{(Fil)}}(t) = A \exp{\left[-t/\left(\tau_{th}\sqrt{1+ZT_{\mathrm{tot}}}\right)\right]} \nonumber  \\
&&  A= \frac{E^2 ZT_{\mathrm{tot}}}{8k_BT^2 G_{\mathrm{th}}^{\mathrm{tot}}} \left(1+ZT_{\mathrm{tot}} \right)^{-1/2}. \nonumber
 \end{eqnarray}
 Therefore for the optimal filter, the plot of $-\ln{I_L^{\mathrm{(Fil)}}(t)}$ vs $t$ is simply a straight line with the slope
$\tau_{th}\sqrt{1+ZT_{\mathrm{tot}}}$.  On the other hand for any random filter $I_L^{\mathrm{(Fil)}}(t)$ does not have simple decaying form but 
further oscillates in time. As a result, the time-averaged current becomes very small.
From this feature of $I_L^{\mathrm{(Fil)}}(t)$
we can identify the pixel for which the designed filter is approximately optimal.
In Fig.~\ref{fig3} we represent the $I_L^{\mathrm{(Fil)}}(t)$ when the filter is optimal (blue curve),
and when the filter is not optimal (orange curve). 

\subsection{Energy resolution of the SF based TED }
Above discussion is valid for a generic TED. In what follows we evaluate
the energy resolution $\Delta E_{\mathrm{opt}}^{(\mathrm{Fil})}$  of a SF based TED,
disregarding the effect of amplifier noise in the measured signal.
We can express the coefficients of the thermoelectric detector as \cite{Tero_2014, Tero_bolometer} 
\begin{eqnarray}
G &=& G_T\int_{-\infty}^\infty dE~ \frac{N_0(E)}{4k_BT\cosh^2\left(\frac{E}{2k_BT}\right)} \label{res1} \\
G_{th} &=& \frac{G_T}{e^2}\int_{-\infty}^\infty dE~ \frac{E^2N_0(E)}{4k_BT^2\cosh^2\left(\frac{E}{2k_BT}\right)} \label{res2} \\
\alpha &=& \frac{PG_T}{2e}\int_{-\infty}^\infty dE~ \frac{EN_z(E)}{4k_BT\cosh^2\left(\frac{E}{2k_BT}\right)}. \label{res3}
\end{eqnarray}
Here $P=\left(G_{\uparrow}-G_{\downarrow}\right)/\left(G_{\uparrow}+G_{\downarrow}\right)$
is the spin polarization, $G_{\sigma}$ is the normal-state conductance for spin $\sigma$,
$N_0(E)=(N_{\uparrow}+N_{\downarrow})/2$ and  $N_z(E)=N_{\uparrow}-N_{\downarrow}$
are the spin-averaged and spin-difference density of states of the superconductor, normalized
to the normal-state density of states, $\nu_F$, at the Fermi level.
Here $N_{\uparrow/\downarrow}=N_S(E\pm h)$ with 
$N_S(E)=\mathrm{Re}\left[ |E+i\Gamma|/\sqrt{(E+i\Gamma)^2-\Delta^2} \right]$,
$h$ is the spin splitting exchange field, and 
$\Gamma \ll \Delta$ describes pair-breaking inside the superconductor. The heat capacity of the absorber
with the volume $\Omega$ of the superconductor is $C_h=\nu_F\Omega e^2G_{th}/G_T$ \cite{Tero_bolometer}. Finally, the
electron-phonon heat conductance is obtained from
\cite{Tero_bolometer, Tero_2017}
\begin{subequations}
\begin{eqnarray}
&& G_{q-ph} = \frac{\Sigma\Omega}{96\zeta(5)k_B^6T^2}\int_{-\infty}^{\infty} dE~E\int_{-\infty}^{\infty}d\omega~
\omega^2|\omega|  \nonumber \\
&&~~~~~~~~~~~ \times L_{E,E+\omega}F_{E,\omega} \label{res4} 
\end{eqnarray}
\begin{eqnarray}
&&\hspace{-7mm} L_{E,E^\prime} = \frac{1}{2}\sum_{\sigma=\uparrow,\downarrow}N_{\sigma}(E)N_{\sigma}(E^\prime) \nonumber \\
&&\hspace{-4mm}~~~~~~~~ \times \left[ 1-\Delta^2/[(E+\sigma h)(E^\prime +\sigma h)] \right]  \label{res5}  
\end{eqnarray}
\begin{eqnarray}
&&\hspace{-7mm} F_{E,\omega} = -\frac{1}{2}\left[ \sinh\left(\frac{\omega}{2k_BT}\right)
\cosh\left(\frac{E}{2k_BT}\right) \right. \nonumber \\
&&\hspace{-4mm}~~~~~~~~~~ \times \left. \cosh\left(\frac{E+\omega}{2k_BT}\right)  \right]^{-1}. \label{res6}
\end{eqnarray}
\end{subequations}
In Eq.~\eqref{res4} $\Sigma$ is the material dependent electron-phonon coupling constant and $\zeta(5)$
is the Riemann zeta function. For $k_BT\ll \Delta -h$, the thermoelectric
coefficients have the analytical estimates \cite{Tero_2014, Tero_bolometer}
\begin{eqnarray}
&&G \approx  G_T\sqrt{2\pi\tilde{\Delta}}\cosh(\tilde{h}) e^{-\tilde{\Delta}}  \label{res7}\\
&&G_{th} \approx  \frac{k_BG_T\Delta}{e^2}\sqrt{\frac{\pi}{2\tilde{\Delta}}} e^{-\tilde{\Delta}}
\left[ e^{\tilde{h}}(\tilde{\Delta}-\tilde{h})^2 \right. \nonumber \\
&&~~~~~~~~ \left. +e^{-\tilde{h}}(\tilde{\Delta}+\tilde{h})^2 \right]  \label{res8}\\
&&\alpha \approx  \frac{PG_T}{e}\sqrt{2\pi\tilde{\Delta}} e^{-\tilde{\Delta}} \left[ \Delta\sinh(\tilde{h})-h\cosh(\tilde{h}) \right]~~~
\label{res9}  \\
&&G_{q-ph} \approx \frac{\Sigma\Omega}{96\zeta(5)}T^4\left[ \cosh(\tilde{h})e^{-\tilde{\Delta}} f_1(\tilde{\Delta}) \right. \nonumber \\
&&~~~~~~~~~~~ \left. + \pi\tilde{\Delta}^5e^{-2\tilde{\Delta}} f_2(\tilde{\Delta})\right],  \label{res10}
\end{eqnarray}
where $\tilde{h}=h/k_BT$ and $\tilde{\Delta}=\Delta/k_BT$. In Eq.~\eqref{res10} the terms $f_1$ and $f_2$
represent the scattering and recombination processes. The functions 
$f_1(x)=\sum_{n=0}^3 C_n/x^n$ and $f_2(x)=\sum_{n=0}^2 B_n/x^n$, where
$C_0=440$, $C_1=-500$, $C_2=1400$, $C_3=-4700$, $B_0=64$, $B_1=144$, $B_2=258$.
The analytical estimate of the $\Delta E_{\mathrm{opt}}^{(\mathrm{Fil})}$ can now be obtained by substituting
Eqs.~\eqref{res7}-\eqref{res10} in Eq.~\eqref{en12}. As at low temperatures the scattering contribution
dominates over recombination in the quasiparticle-phonon heat conductance,
we neglect the recombination process and assume $f_1(x)\approx 400$ for $k_BT\lesssim 0.1\Delta$
to obtain a simplified analytical estimate of the energy resolution $\Delta E_{\mathrm{opt}}^{(\mathrm{Fil})}$ of the SF based TED.
\begin{eqnarray}
&&\Delta E_{\mathrm{opt}}^{(\mathrm{Fil})} \approx  \left[\sqrt{4\nu_{\mathrm{F}}\Omega k_B^3T^3 \chi /ZT}\right]\left(1+ZT\right)^{1/4}, \label{res11} \\
&&\chi = \left(2\pi \tilde{\Delta}\right)^{1/2}e^{-\tilde{\Delta}}\left[ \left(\tilde{\Delta}^2+\tilde{h}^2\right)\cosh\left(\tilde{h}\right) \right. 
~~~\nonumber \\
&&~~~~~~\left.-2\tilde{\Delta}\tilde{h} \sinh\left(\tilde{h}\right) \right],  \\
&& ZT \approx  \frac{P^2}{1-P^2+\frac{\tilde{\Delta}^2+Z_{\mathrm{spur}}\cosh^2\left(\tilde{h}\right)}{\left[\tilde{h}\cosh\left(\tilde{h}\right)  
 -\tilde{\Delta}\sinh\left(\tilde{h}\right)\right]^2}}, \\
&& Z_{\mathrm{spur}} = \frac{e^2\Sigma\Omega\Delta^3}{G_Tk_B^5} \tilde{\Delta}^{-4} \frac{220}{96\zeta(5)}.
\end{eqnarray}
\begin{figure}[h]
\includegraphics[width=8cm]{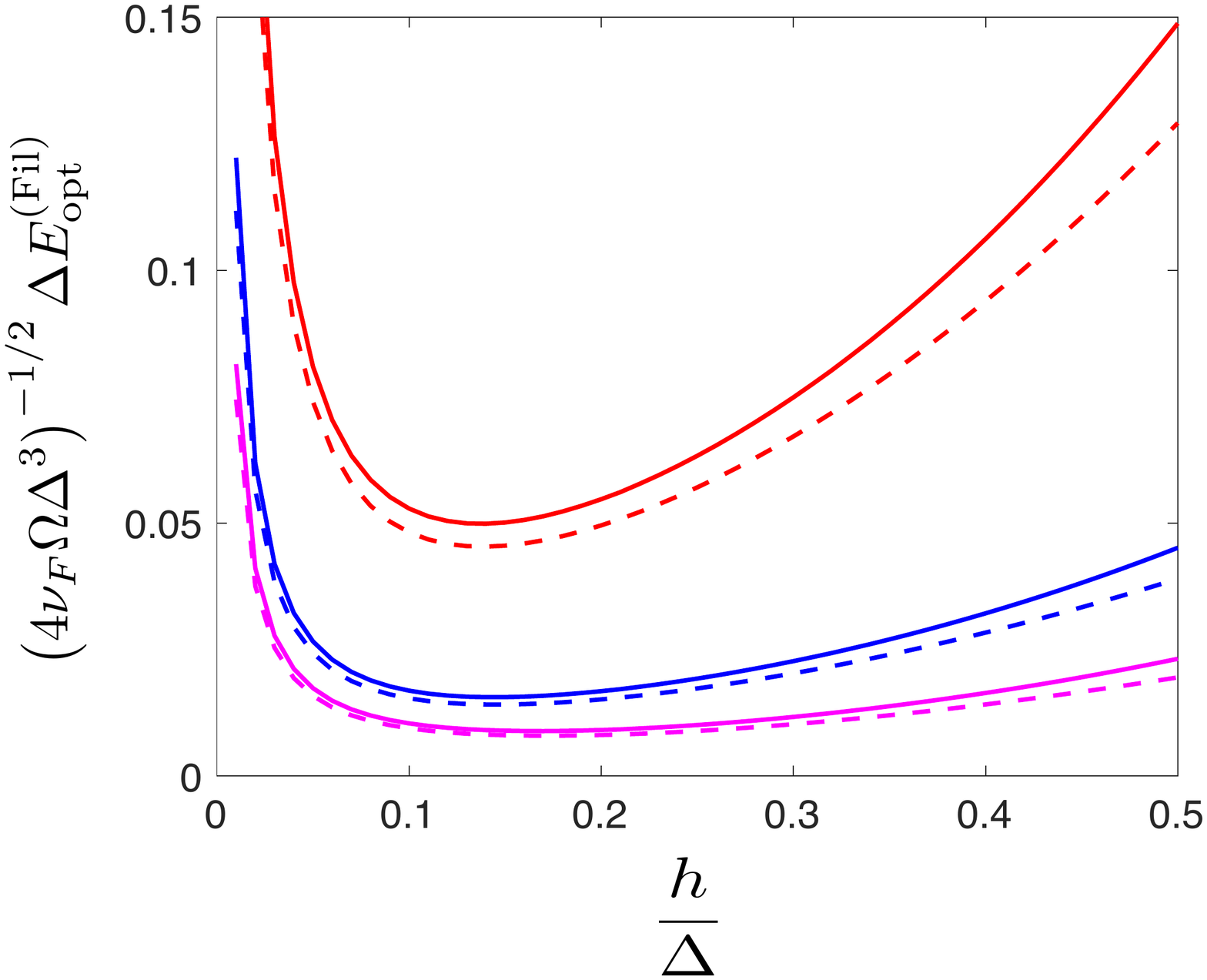}
\caption{\label{en1}  Energy resolution after optimal filtering as a function of exchange field  
for $\Gamma=10^{-4}\Delta$, $k_BT=0.1\Delta$ and $G_T=5\times 10^{-4}~e^2\Sigma\Omega\Delta^3/k_B^5$,
where red, blue and magenta lines respectively represent the plots for $P=0.2$,
$P=0.6$ and $P=0.9$. The solid lines are obtained numerically whereas the dashed lines
are the analytical estimates from Eq.~\eqref{res11} for the corresponding situations.  }
\end{figure}
\begin{figure}[h]
\includegraphics[width=8cm]{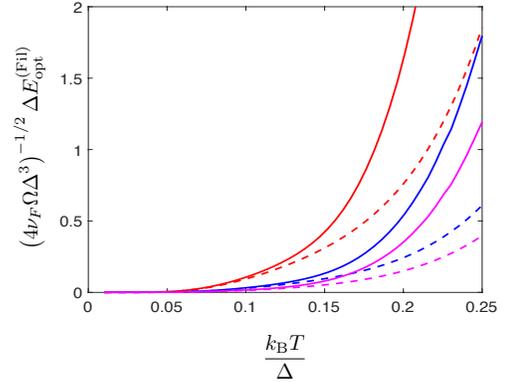}
\caption{\label{en2} Energy resolution after optimal filtering as a function of temperature 
for $\Gamma=10^{-4}\Delta$, $h=0.4\Delta$ and $G_T=5\times 10^{-4}~e^2\Sigma\Omega\Delta^3/k_B^5$,
where red, blue and magenta lines respectively represent the plots for $P=0.2$,
$P=0.6$ and $P=0.9$. The solid lines are obtained numerically whereas the dashed lines
are the analytical estimates  from Eq.~\eqref{res11} for the corresponding situations.}
\end{figure}
In the following, we use the formulas of Eqs.~\eqref{en12}-\eqref{res6} to find
$\Delta E_{\mathrm{opt}}^{(\mathrm{Fil})}$ as a function of the exchange field and the bath temperature of the SF based TED. 
We also compare the numerical results with the analytical estimate in Eq.~\eqref{res11}. 
We consider an Al absorber of volume $\Omega=10^{-19}~ \mathrm{m}^3$ and the superconducting
critical temperature at zero exchange field $T_c=1.2 $ K \cite{Matthias}, $\nu_{\mathrm{F}}=10^{47} ~\mathrm{J}^{-1}\mathrm{m}^{-3}$ 
 and $G_T=5\times 10^{-4} e^2\Sigma \Omega\Delta^3/k_B^5 \sim 25~ \mu{\mathrm{S}}$ \cite{Tero_bolometer}.
 With these choices, we get an overall scaling factor
 $\sqrt{4\nu_F\Omega\Delta^3}=20$ meV. This factor is used in
 Figs.~\ref{en1} and \ref{en2} as the unit energy resolution.  
We thus find that the optimal energy resolution at $k_B T=0.1\Delta$,
corresponding to $T=$200 mK for Al, can be below 1 meV. This 
corresponds to single-photon resolution at frequencies $f =2\pi \Delta
E_{\rm opt}^{\rm Fil}/\hbar$ above 240 GHz. Figure \ref{en2} shows the
corresponding temperature dependence of the optimal energy
resolution (solid lines). The analytical estimate in Eq.~\eqref{res11}
fits the numerics up to $k_B T \lesssim 0.1 \Delta$. Above this the
quasiparticle-phonon recombination process starts affecting the results.

Let us again discuss the added noise in current amplification. A good
cryogenic SQUID amplifier can reach $S_A \sim 0.3$ (fA)$^2$/Hz \cite{beev13}. With
the above-chosen tunnel conductance and superconducting gap $\Delta$, this
would translate into $ZT/ZT_A  \approx 2 \times 10^{-4} \times
(1+ZT) G_T \Delta/(G k_B T)$. This becomes of the order of unity or larger for
$k_B T \lesssim 0.1 \Delta$. Below that temperature it would be
advantageous to either measure the voltage instead of the current, or
use higher-conductance junctions. Contrary to the noise equivalent
power in bolometers
\cite{Tero_bolometer}, this does not deteriorate the energy
resolution. However, increasing the contact transparency may be
challenging especially with Al/EuS based spin-filter junctions.

\section{Conclusions}
In this work we present the first full noise analysis of a generic
thermoelectric detector TED, especially including the possibility of a
high thermoelectric figure of merit. In particular, we show that TEDs
based on superconductor-ferromagnet systems may rival the best
transition edge sensor (TES) -type calorimeters, reaching wide-band energy resolution
below 1 meV (with unit quantum efficiency). Hence such SF TEDs present
a viable alternative for TES devices especially in the case of large
arrays where the lack of required probe power leads to reduced heating
and simplified design of the detectors. 

We thank Ilari Maasilta for discussions. This project was supported by
the Academy of Finland Key Funding project (Project No. 305256), and the Center of Excellence program (Project No. 284594).

\bibliography{Reference}

\end{document}